\documentclass{PoS}

\title{The 3rd Catalog of AGN Detected by the Fermi LAT}

\ShortTitle{The 3LAC Catalog}

\author{\speaker{Dario Gasparrini}\\\
        Agenzia Spaziale Italiana (ASI) Science Data Center, I-00133, Roma, Italy\\
        Istituto Nazionale di Fisica Nucleare, sezione di Perugia, I-06123, Perugia, Italy\\
        E-mail: \email{gasparrini@asdc.asi.it}}

\author{Benoit Lott\\
        \\Centre d'\'Etudes Nucl\'eaires de Bordeaux Gradignan, IN2P3/CNRS, Universit\'e Bordeaux 1, BP120, F-33175 Gradignan Cedex, France\\
        E-mail: \email{lott@cenbg.in2p3.fr}}

\author{Sara Cutini\\
        Agenzia Spaziale Italiana (ASI) Science Data Center, I-00133, Roma, Italy\\
        Istituto Nazionale di Fisica Nucleare, sezione di Perugia, I-06123, Perugia, Italy\\
        E-mail: \email{cutini@asdc.asi.it}}

\author{Stefano Ciprini\\
        Agenzia Spaziale Italiana (ASI) Science Data Center, I-00133, Roma, Italy\\
        Istituto Nazionale di Fisica Nucleare, sezione di Perugia, I-06123, Perugia, Italy\\
        E-mail: \email{ciprini@asdc.asi.it}}

\author{Elisabetta  Cavazzuti\\
        Agenzia Spaziale Italiana (ASI) Science Data Center, I-00133, Roma, Italy\\
        E-mail: \email{cavazzuti@asdc.asi.it}}

\author{on behalf of the {\it Fermi}-LAT Collaboration}

\abstract{The third catalog of active galactic nuclei (AGNs) detected by the Fermi Large Area Telescope (3LAC) is presented. It is based on the third catalog (3FGL,\cite{3FGL}) of sources detected with a test statistic greater than 25, using the first 4 years of data. The 3LAC includes 1591 AGNs located at high ($|b|>10^\circ$) Galactic latitudes (with 28 duplicate associations, thus corresponding to 1563 gamma-ray sources among 2192 sources in the 3FGL catalog), providing $71\%$ more sources with respect to the 2FGL. Various properties, such as gamma-ray fluxes and photon power law spectral indices, redshifts, gamma-ray luminosities, variability, and their correlations are presented and discussed for the different blazar and non-blazar classes.}

\FullConference{The 34th International Cosmic Ray Conference,\\
		30 July- 6 August, 2015\\
		The Hague, The Netherlands}

\begin{document}

\section{Introduction}

Since its launch in 2008, the {\it Fermi}-LAT has revolutionised our knowledge of the gamma-ray sky above 100 MeV.
 Its unique combination of high sensitivity, wide field of view, large energy range, and nominal sky-survey operating mode  has enabled a complete mapping and continuous monitoring of the gamma-ray sky to an unprecedented level.   
Several catalogs and sources lists, both general and specialized (AGNs, pulsars, supernova remnants, pulsar wind nebulae, gamma-ray bursts, very-high-energy candidates) have already been produced. These constitute important resources to the astronomical community. The successive AGN lists and catalogs, LBAS (LAT Bright AGN Sample\cite{LBAS}), 1LAC \cite{1LAC} and  2LAC \cite{2LAC,2LACerr}, first and second LAT AGN catalogs respectively, have triggered numerous population studies, provided suitable samples, e.g., to probe the Extragalactic Background Light, offered suitable target lists to investigate the dichotomy between  gamma-ray loud and gamma-ray quiet blazars at other wavelengths, and serve as references for works on individual sources. 

Here we present the  third catalog of AGNs detected by the {\it Fermi}-LAT  after four years of operation (3LAC \cite{3LAC}).  It makes use of the results of the 3FGL catalog \cite{3FGL} which includes 3033 sources with  a Test Statistic\footnote{We use the Test Statistic $TS = 2\Delta\log L$ for quantifying how significantly a source emerges from the background, comparing
the likelihood function $L$ with and without that source.}  ($TS$) greater than 25.   Among them,  2192 sources are detected at $|b|>10^\circ$ where $b$ is the Galactic latitude. Among these 2192, 1563 (71\%) are associated with high confidence with 1591 AGNs, which constitute the 3LAC.  The 3LAC represents a sizeable improvement over the 2LAC as it includes 71\% more sources (1591 vs. 929\footnote{see \cite{2LACerr} for a 2LAC Erratum. The corrected 2LAC full and clean samples include 929 and 827 sources, respectively.}) with an updated data analysis.  We will briefly review the association/classification procedure and then present the properties of the 3LAC sample. We will refer the reader to the 3FGL paper for technical details regarding the LAT data analysis.     

\section{Source association and classification}

We use the same two association methods used in the 2LAC: the Bayesian Association Method and Likelihood Ratio Method. We apply the Bayesian Association Method \cite{1FGL} to catalogs of sources that were already classified and/or characterized. These catalogs  come from specific instruments providing information on the spectrum and/or broadband emission. If a catalog reports an AGN classification, that is used.  Otherwise the classification is made according to the criteria described below.
This powerful method, however, cannot deal with large all-sky surveys, which would be too computationally intensive.  Therefore, to broaden the possibility of associating a candidate AGN knowing its broadband emission characteristic we added the Likelihood Ratio Method \cite{2LAC}.  This method can handle large uniform all-sky surveys and take the source space-density distribution into account. In the case of general radio or X-ray surveys,  including AGN and non-AGN sources,  the classification procedure is the same as for the Bayesian Association Method.

The adopted threshold for the association probability is 0.80 in either method. This value represents a compromise between association efficiency and purity. 

The fraction of sources associated by both methods is 71\% (1150/1591), 379 and 62 sources being solely associated with the Bayesian and LR methods respectively.  

 The overall false-positive rate is 1.9\%. The estimated number of false positives among the 571 sources not previously detected in 2FGL and previous LAT catalogs is 12.0 (2.1\%). 
 
 To define the criteria that a source must fulfill to be considered as an AGN, the ingredients are primarily the optical spectrum and to a lesser extent other characteristics such as radio loudness, flat/steep radio spectrum between 1.4 GHz and 5 GHz, broadband emission, variability, and polarisation.

Besides the traditional optical classification (FSRQ, BL Lac, etc.) based on the strength of emission lines (only possible when a good quality spectrum is available), an SED classification based on the position of the synchrotron peak into low-, intermediate-, and high-synchrotron peaked sources was made using a human-controlled fit for each source individually. In addition, sources with some properties typical of blazars (blazars of uncertain/transitional type, flat radio spectrum typical two-humped spectral energy distribution) were labelled as blazar candidates of unknown type (BCUs). In this new fit, some sources changed SED classification with respect to the2LAC, where the procedure was run in a fully automatic way.

 \subsection{\label{sec:census} Census}

The 3LAC includes 1591 objects with 467 FSRQs, 632 BL~Lacs, 460 BCUs and 32 non-blazar AGNs.  The latter comprise 12 FR I, 3 FR II, 8 steep spectrum radio sources and 5 Narrow Line Radio Loud Seyfert 1 galaxies.
 A total of 1563 gamma-ray sources have been associated with radio-loud AGNs among 2192 $|b|>10^\circ$ 3FGL sources, corresponding to an overall association fraction of 72\%.

 As in previous LAC catalog versions, we define a Clean Sample by excluding sources that have any of the 3FGL analysis flags set and excluding the multiple-association sources. 
 It includes 1444 objects with 414 FSRQs, 604 BL~Lacs, 402 BCUs and 24 non-blazar AGNs.  

A comparison of the results inferred from the 3LAC and 2LAC enables the following observations:
\begin{itemize}
\item The 3LAC Clean Sample includes 619  more sources than the 2LAC Clean Sample, i.e., a  75\% increase.  Of these, 477 sources are new (81 FSRQs, 146 BL~Lacs, 240 blazars of  unknown type, 10 non-blazar objects); the other sources were present in previous {\it Fermi} catalogs but not included in Clean Samples for various reasons (e.g., the corresponding  gamma-ray sources were not associated with AGNs, had more than one counterpart or were flagged in the analysis).  
\item The fraction of blazars of unknown type (BCU) has increased notably between the two catalogs (from 20\% to 28\%). The number of these sources in the 3LAC Clean Sample has increased by more than a factor of 2.5  relative to that in the 2LAC Clean Sample, being almost equal to the number of FSRQs. This increase is mainly due to the lower probability of having a published high-quality spectrum available for these fainter sources because of the lack of  optical observing programs. 
\item The relative increase in BCUs drives a drop in the proportions of FSRQs and BL~Lacs, which only represent 29\% and 41\% of the 3LAC Clean Sample respectively  (38\% and 48\% for 2LAC). The relative increase in the number of sources with respect to 2LAC is 34\% and 42\% for FSRQs and BL~Lacs respectively. 
\item Out of 827 sources in the 2LAC Clean Sample, a total of 69 are missing in the 3LAC Clean Sample (42 in the full sample), some of them probably due to variability effects.  A few others are present in 3FGL but with shifted positions, ruling out their association with their former counterparts.
\end{itemize}

\subsection{\label{sec:lowlat}Low-Latitude AGNs}

Because of the intrinsic incompleteness of the counterpart  catalogs in this sky area ($|b|<10^\circ$), these sources are treated separately and are not included in the 3LAC. 
We report associations for 182 blazars (75\% more than in 2LAC) located at $|b|<10^\circ$ : 24 FSRQs, 30 BL Lacs, 125 BCUs and 3 non-blazar AGNs. Extrapolating from the number of high-latitude sources and assuming the same sensitivity, about 340 sources would be expected in this area.   The discrepancy between expected and actual source numbers stems from the dual effect of a higher detection threshold due to a higher Galactic diffuse emission background and a higher incompleteness of the counterpart  catalogs for this area.

\section{Properties of 3LAC Sources}

\subsection{\label{sec:flux}Flux and Photon Spectral Index}

The newly-detected FSRQs are slightly softer than the 2LAC ones (2.53$\pm$0.03 vs. 2.41$\pm$0.01), indicating that the LAT  gradually detects more lower energy-peaked blazars. In contrast,  there is no significant spectral difference between the two sets of BL~Lacs. For BCUs, the distribution of the new sources extends further out on the high-index end ($\Gamma>$2.4), where the overlap with the BL~Lac distribution becomes very small.   The corresponding sources seem likely to be FSRQs.  The strong anticorrelation between the  position of the synchrotron peak  $\nu^S_{peak}$ and the photon spectral index for FSRQs and BL~Lacs  is confirmed. A similar trend is actually observed for BCUs supporting the idea that BCUs with low  $\nu^S_{peak}$ and  high $\Gamma$ are likely FSRQs, while the rest would  mostly be BL~Lacs.

\subsection{\label{sec:z}Redshift}

The redshift  distributions are fairly similar, although the newly detected  FSRQs are located at slightly higher redshift than the 2LAC ones ($\langle z \rangle$=1.33$\pm$0.08 vs. 1.17$\pm$0.03). The maximum redshift for an FSRQ is still 3.1 (four FSRQs have 2.94 $< z <$3.1) and has not changed since the 1LAC. This trend allowed the conclusion that the number density of FSRQs grows dramatically up to redshift $\simeq$0.5-2.0 and declines thereafter \cite{Aje12}.  

The redshift distribution of new BL~Lacs is somewhat narrower than that of the 2LAC sources, with a maximum near z=0.3. The redshift distributions gradually spread out to higher redshifts when moving from  HSP-BL Lacs to LSP-BL Lacs, a feature already seen in 2LAC. However, the HSP distribution extends to higher redshifts relative to 2LAC, with four  HSPs having measured redshifts greater than 1  and one having a redshift greater than 2. Three of these four HSPs were already included in 2LAC but either lacked measured redshifts or were classified differently.  

Among BL~Lacs, 309 have a measured redshift, while 295 do not.
The fraction of BL~Lacs without redshift is
55\%, 61\% and 40\% for LSPs, ISPs and HSPs respectively. However, \cite{Sha13} have provided  redshift constraints for 134 2LAC BL~Lacs: upper limits from the absence of Ly$\alpha$ absorption for all of them and lower limits from non-detection of the host galaxy or from intervening absorption line systems for a subset of 57 objects. It was noted by these authors that the average lower limit exceeded the average measured redshift for BL~Lacs, indicating that the measured redshifts are biased low. 
The redshift ranges are very similar for the different subclasses and all cluster at high redshifts, with a median around $z$=1.2. This is in good agreement with the predictions of \cite{Gio13}.

\subsection{Luminosity}

Sources with high gamma-ray luminosity (mostly FSRQs) are found to have softer spectra on the average than low-luminosity sources (mostly BL Lacs). This correlation has been widely discussed in the context of the ``blazar divide'' or ``blazar sequence'' .  Because of the redshift bias mentioned above, the HSPs with both limits are more luminous on average than those with measured redshifts, thus populating a previously scarcely occupied area in the $L_\gamma$-$\Gamma$ diagram. This observation has profound consequences for the blazar sequence.  Note that \cite{Aje14} found a small but significant correlation between gamma-ray luminosity and spectral index when including the redshift constraints from \cite{Sha13}.

\subsection{Spectral curvature}

First observed for 3C~454.3 \cite{Abdo_3C} early in the {\it Fermi} mission, a significant curvature in the energy spectra of many bright FSRQs and some bright LSP-/ISP-BL~Lacs is now a well-established feature \cite{1LAC}. The break energy obtained from a broken power-law fit has been found to be remarkably constant as a function of flux, at least for 3C~454.3. Several explanations  have been proposed to account for this feature,  including $\gamma\gamma$ attenuation from He\,{\sc ii} line photons \cite{Pou10},
intrinsic electron spectral breaks \cite{Abdo_3C}, Ly$\alpha$ scattering
\cite{Abdo_3C_10}, Klein-Nishina effects taking place when jet electrons scatter BLR radiation in a near-equipartition approach and hybrid scattering \cite{Fin10}. The level of curvature has been observed to diminish during some flares.   

A total of 91 FSRQs (57 in 2LAC), 32 BL~Lacs (12 in 2LAC) and 8 BCUs show significant curvature at a confidence level $>$99\%. 

\subsection{\label{sec:var} Variability}

Variability is a key feature of blazars. The 3FGL monthly light curves provide a baseline reference against which other analyses can be cross-checked and enable cross-correlation studies with data obtained at other wavelengths. Although variability at essentially all time scales has been observed in blazars, the monthly binning represents a trade off between a shorter binning needed to resolve flares in bright sources and a longer binning required to detect faint sources. Even so, only 15 sources are detected in all 48 bins with monthly significance $TS>$25, while this number becomes 46 if a relaxed condition $TS>$4 is required.

The features already reported in 2LAC are again visible, with a large fraction of FSRQs found to be variable (69\%), with the fraction for BL~Lacs much lower on average (23\%) and with a steadily decreasing trend as $\nu^S_{peak}$ rises  (39\%, 23\%, 15\% for LSPs, ISPs and HSPs respectively). These fractions are quite similar to those reported in 2LAC, despite the larger population and longer time span of the light curves. A similar trend between variability index and $\nu^S_{peak}$  is observed for blazars of unknown type  with 21\% of them found to be variable.

\section{Multiwavelength Properties of 3LAC Sources}

It was shown in 2LAC that the LAT-detected blazars display on average larger
radio fluxes than non-detected blazars and that they are all bright in the optical.  We focus on the connection with the two neighboring  bands, namely the hard X-rays and the VHE bands. 


A total of 85 3LAC sources are in common with the {\it Swift} BAT 70-month survey \cite{BATcatalog} in the 14-195 keV band performed between December 2004
and September 2010  (there were 47 in 2LAC). These 85 sources include 34 FSRQs with an average redshift of 1.37$\pm$0.15.  Only 9 BAT FSRQs are missing from 3LAC. Out of 37 BAT BL~Lacs, 30 have now been detected with the LAT.  

It is also worth noting that 96 3LAC sources (5 Radio Galaxies, 53 FSRQs, 33 BL~Lacs, 4 BCUs, 1 NLSy1) are present in the V38 INTEGRAL source catalog\footnote{http://www.isdc.unige.ch/integral/science/catalogue}, which includes 540 AGNs located at $|b|>10^\circ$. 


At the time of writing, some 56 AGNs which have been detected at TeV energies are listed in TeVCat \footnote{http://tevcat.uchicago.edu}. Among them,  55 are present in 3FGL, which is a remarkable result underscoring the level of synergy that has now been achieved between the high-energy and VHE domains. 
Only 28 out of the 55 3FGL sources are seen to be variable in the LAT energy range at a significance greater than 99\%.

\section{Discussion}

\subsection{Gamma-ray detected versus non-detected blazars}

The   blazars detected in gamma rays after 4 years of LAT operation represent a sizeable fraction of the whole population of known blazars as listed in BZCAT.  BZCAT represents an exhaustive list of sources ever classified as blazars but is by no means complete. Although a comparison between the gamma-ray detected and non-detected blazars within that sample has no strong statistical significance in terms of relative weights, it is nevertheless useful to look for general trends.

The overall LAT-detected fraction is 24\% (409/1707) for FSRQs, 44\% (543/1221) for BL Lacs and  27\%  (59/221) for BCUs.
The redshift distributions are quite similar for the two subsets although the distribution for the blazars  undetected  by the LAT extends to significantly higher redshifts. The  highest-redshift BZCAT sources  (56 have z$>$3.1 reaching z=5.47) are still eluding detection by the LAT.

The similarity of the redshift or flux distributions (radio, optical and X-ray band) between the detected and non-detected sets of blazars supports the conjecture that they belong to the same population of sources intermittently shining in gamma rays.

\section{Conclusions}

We have presented the third catalog of LAT-detected AGNs (3LAC), based on 48 months of LAT data.  This is an improvement over the 1LAC (11 months of data) and 2LAC (24 months of data) also in terms of data quality and analysis methods. Key results from the 3LAC sample include: 

1. An increase of 71\% in the number of blazars relative to 2LAC is  associated with the two-fold increase in exposure. 

2. A significant increase of  the non-blazar population  is found with respect to previous catalogs. {The new sources include:} two FRIIs, three FRIs  and four SSRQs. However other sources reported in previous catalogs are now missing.
 
3. A large fraction ($>75$\%) of {\sl Swift} hard X-ray BAT-detected blazars and all but one TeV-detected AGNs have now been detected by the ${\it Fermi}$-LAT.  

4. The most distant 3LAC blazar is the same as in 1LAC and 2LAC,  PKS~0537$-$286 lying  at z=3.1.  Although 50\% of the BL~Lacs still do not have measured redshifts, upper limits have recently been obtained for 134 2LAC sources and lower limits as well for 57 of them. These constraints indicate that the measured redshifts are biased low for BL~Lacs. Using the luminosities derived from these constraints, the sources populate a previously scarcely occupied area in the $L_\gamma$-$\Gamma$ diagram, somewhat undermining the picture of the blazar sequence. 

5. Along the same lines, a few rare outliers (four high-luminosity HSP BL~Lacs and two HSP FSRQs) are included in the 3LAC, while they were missing in 2LAC.   The high-luminosity HSP-BL~Lacs exhibit  Compton dominance values similar to the bulk of that class.  
 
6. The main properties of blazars previously reported in 1LAC and 2LAC are confirmed. The average photon index, gamma-ray luminosity, flux variability, spectral curvature monotonically evolve from FSRQs to HSP BL Lacs with  LSP- and ISP-BL~Lacs showing intermediate behavior.  

7. No strong differences in the radio, optical and X-ray flux distributions are observed between gamma-ray detected and non-detected BZCAT blazars. This is a clue that essentially all known blazars could eventually shine in gamma rays at LAT-detection levels. A larger fraction (44\%) of the known BL~Lacs than FSRQs (24\%) has been detected so far.  The duty cycle of FSRQs appears to be longer than four years.

The 3LAC catalog is intended to serve as a valuable resource for a better understanding of the gamma-ray loud AGNs. The next LAT AGN catalog will benefit from the improved Pass 8 data selection and IRFs \cite{Atw13}. Pass 8 is the result of a comprehensive revision of the entire event-level analysis,  based on the experience gained in the prime phase of the mission.  The gain in effective area  at the low end of the LAT energy range will be particularly notable. The 4LAC catalog is thus  expected to include a non-incremental number of new, especially soft-spectrum AGNs.

\acknowledgments { The \textit{Fermi}-LAT Collaboration acknowledges support for LAT development, operation and data analysis from NASA and DOE (United States), CEA/Irfu and IN2P3/CNRS (France), ASI and INFN (Italy), MEXT, KEK, and JAXA (Japan), and the K.A.~Wallenberg Foundation, the Swedish Research Council and the National Space Board (Sweden). Science analysis support in the operations phase from INAF (Italy) and CNES (France) is also gratefully acknowledged.

}

\bibliographystyle{JHEP}
\bibliography{3lac}


\end{document}